\documentclass{svproc}
\usepackage{url}

\usepackage{graphicx}
\usepackage{amsmath}
\usepackage{amssymb,amsfonts}
\usepackage{algorithm}
\usepackage{algorithmic}

\begin{document}
\mainmatter              
\title{Real-time and Zero-footprint Bag of Synthetic Syllables Algorithm for E-mail Spam Detection Using Subject Line and Short Text Fields}
\titlerunning{Bag of Synthetic Syllables Algorithm}  
%
\author{Stanislav Selitskiy}
\authorrunning{Stanislav Selitskiy} 
%
\tocauthor{Stanislav Selitskiy}
\institute{Earthlink Internet, Atlanta GA 30328, USA,\\
\email{stanislav.selitskiy@elnk.com}
}

\maketitle              

\begin{abstract}
Contemporary e-mail services have high availability expectations from the customers and are resource-strained because of the high-volume throughput and spam attacks. Deep Machine Learning architectures, which are resource hungry and require off-line processing due to the long processing times, are not acceptable at the front line filters. On the other hand, the bulk of the incoming spam is not sophisticated enough to bypass even the simplest algorithms. While the small fraction of the intelligent, highly mutable spam can be detected only by the deep architectures, the stress on them can be unloaded by the simple near real-time and near zero-footprint algorithms such as the Bag of Synthetic Syllables algorithm applied to the short texts of the e-mail subject lines and other short text fields.
The proposed algorithm creates a circa 200 sparse dimensional hash or vector for each e-mail subject line that can be compared for the cosine or euclidean proximity distance to find similarities to the known spammy subjects. The algorithm does not require any persistent storage, dictionaries, additional hardware upgrades or software packages. The performance of the algorithm is presented on the one day of the real SMTP traffic.
\keywords{Spam detection, bag of features, short text, e-mail subject, online training, proximity metrics}
\end{abstract}

\section{Introduction}
\begin{figure}
\begin{minipage}[b]{0.8\linewidth}
  \centering
  \centerline{
  \includegraphics[width=0.415\linewidth]{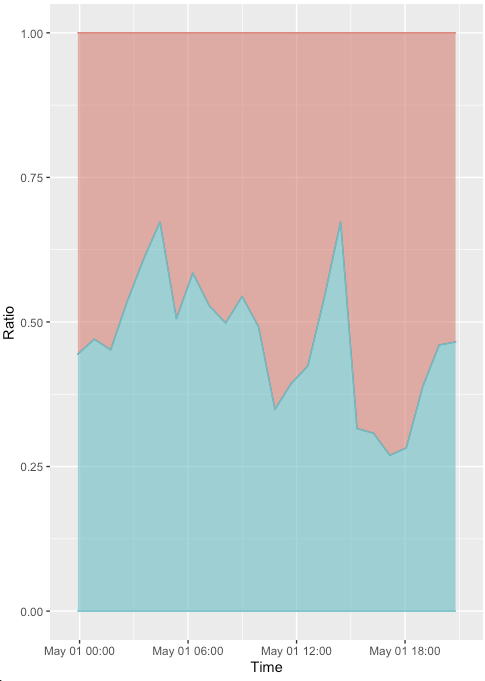}
  \includegraphics[width=0.385\linewidth]{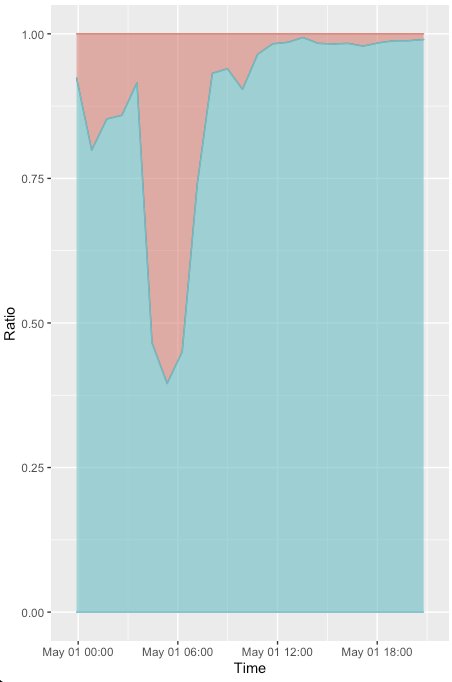}
  }
\end{minipage}
\caption{Left: Spam (bottom) to No Spam (top) ratio (after IP filtering). Right: dumb (bottom) to intelligent (top) spam ratio. 24 hours snapshot.}
\label{fig:spam_no}
\end{figure}

Level of the spam e-mail traffic coming through the Simple Mail Transfer Protocol (SMTP) \cite{Klensin2008Oct}, circa 90\% before or 50\% after IP filtering, makes it effectively nonfunctional without filtering neither for users nor economically sound for the Internet Service Provider (ISP) companies. 
The majority of the practical anti-spam solutions rely on crowd-sourcing and, partially, expert analysis of the spam-attracting honey-pot accounts to extract signatures from the spam message example. Such signatures include IP addresses, handshake and source domains, header domains, subject and other text headers, body text, URLs, and attachments. Filtering on such signatures is usually effective, in term of accuracy and speed, against the non-sophisticated spam comprising about 90\% of all spam traffic Figure~\ref{fig:spam_no}. However, such signatures become available only a few hours after the spam attack with unknown previously signatures starts. Also, keeping and searching databases of spam signatures requires either significant computing and storage resources on-site or paid subscription to the spam-filtering providers.

Intelligent spammers are aware of these limitations and exploit them by running distributed, short-lived, intense campaigns Figure~\ref{fig:spam_no}.right, rotating spam signatures, monitoring the anti-spam algorithms' effectiveness via so-called canary accounts, and crafting unique spam messages individually tailored for each recipient. Deep Learning (DL) algorithms can detect sentiment and semantic of the intelligent spam full-body texts \cite{10.1145/3014812.3014815,Jain2019}. 

However, DL algorithms require significantly more resources and have longer processing time than simpler algorithms. Although SMTP standards allow plenty of time for a message to be delivered to the recipient, contemporary e-mail users expect near real-time message delivery. Therefore, slow and expensive DL algorithms tend to be used on the last line of the defence for messages with the unclear verdict. Another indirect impact of the full-body message scan has game-theoretic consequences - it increases the incoming spam messages' size up to the maximal limits because spammers try to overwhelm spam filters. Therefore, for employing full-body analysis, ISP should be prepared resource-wise to handle the shifting traffic's structure and volume.

The behavioural-based algorithms use the simplified feature space proximity analysis for the subject line and other short text headers to fill the apparent gap between the static signature-based algorithms and the DL full-body semantic and sentiment analysis algorithms. Bag of Words algorithms is a popular choice for such analysis \cite{Zhang2010}. They use from the few hundred to few thousand-dimensional spaces of the frequency vocabularies and various distance and class boundary algorithms such as cosine and Euclidean distances, or Support Vector Machines (SVM) and Artificial Neural Networks (ANN) regression algorithms. However, Bag of Words models require text pre-processing and database infrastructure that consume time and hardware resources.

The presented Bag of Synthetic Syllables (BoSS) algorithm is self-contained, has a straightforward fast-computing logic, does not require any external resources and introduces minimal CPU or memory-wise overhead. The BoSS algorithm can be viewed as related to n-gram algorithms with custom 2-gram and 1-gram mix that creates enough dimensional space to handle short texts, still maintaining low processing requirements to find morphological or stochastic variation neighbourhoods \cite{BojanowskiGJM16,Sureka2010}.

Machine Learning concepts have been efficiently used for detection of abnormal patterns \cite{Nyah_SAI,Nyah_IS} and estimation of brain development \cite{Jakaite2010a,Jakaite_EEG_CBMS,Schetinin_EEG_CBMS,Jakaite_CMMM,Schetinin2012_ESWA_EEG}, trauma severity estimation \cite{eswa_ntdb_2013,Jakaite_Balt} and survival prediction \cite{eswa_ntdb_2013,AIM2018,ijmi2018}, collision avoidance at Heathrow \cite{icae2018}, and early detection of bone pathologies \cite{Mukti2019,Jakaite2021}. 

The paper is organized as follows. Section~\ref{sec:al} describe the BoSS algorithm in detail. Section \ref{sec:ex} describes the experimental setting and results. Section \ref{sec:disc} draws practical conclusions from the results and states directions of the research of not yet answered questions.

\section{Bag of Synthetic Syllables algorithm}
\label{sec:al}
The algorithm expects the English character set ASCII (American Standard Code for Information Interchange) string of the $1$ kB length (though precise internet message header size is $988$ symbols \cite{Resnick2008Oct}). Any symbol or symbol sequence that does not belong to `a'-to-`z' or `A'-to-`Z' intervals are considered as between the words delimiters. Interval `A'-to-`Z' is converted to the lower case `a'-to-`z'. Out of the $26$ symbols, $6$ (`a', `i', `u', `e', `o', `y') are considered vowels, and the rest $20$ - consonants. Synthetic syllables are constructed in the Japanese morae style - $2$-symbol syllables start with a consonant followed by a vowel. If two vowels follow each other, then two $1$-symbol vowel `syllables' are created. If two consonants follow each other, then one $1$-symbol consonant `syllable' is created. 

This synthetic syllabification schema differs from the native English or other languages with alphabetic writing systems texts on which this algorithm may be applied to. For example, the single syllable word `tree' under this schema will be broken into three synthetic syllables: `t', `re', and `e'. Such an approach allows keeping controlled compact dimensionality of the feature space and fast mapping into it. 

The input short text string then can be represented in the $20 \times 7 + 6 = 146$ dimensional space $\mathcal{S_{BoSS}} \subset \mathbb{I}^{146} = span({\textbf{s}_1 \dots \textbf{s}_{146}})$, where $\textbf{s}_i$ is a basis syllable vector. Similarly, it can be interpreted as a $146$ bin syllable frequency histogram. 

\begin{algorithm}
\caption{The Bag of Synthetic Syllables hash building function}
\label{alg:algorithm1}
\textbf{Input}: Short text string buffer $str$\\
\textbf{Parameters}: BoSS hash length $bss\_len=189$, high (consonant) register lengths $hreg\_len=27$\\
\textbf{Output}: BoSS $hash$\\
\begin{algorithmic}[1] 
\STATE $memset(hash, `0\textrm', bss\_len)$
\STATE $str \gets tolower(str)$
\STATE $state \gets$ `out~of~syllable'
\FORALL {symbols $s_i$, $i \in \{1, \dots |str|\}$}
\IF{$state =$ `out~of~syllable'}

\IF{$s_i \in \{a, \dots z\}$}
\IF{$s_i \in v=\{a, i, u, e, o, y\}$}
\STATE $hreg \gets j$, where $v_j = s_i$, 
\STATE $j \in \{1, \dots 6\}$
\STATE $++hash[hreg*hreg\_len]$
\STATE $state \gets$ `out~of~syllable'
\ELSE
\STATE $lreg \gets s_i - `a\textrm'$
\STATE $state \gets$ `in~syllable'
\ENDIF
\ENDIF

\ELSE

\IF{$s_i \in \{a, \dots z\}$}

\IF{$s_i \in v=\{a, i, u, e, o, y\}$}
\STATE $hreg \gets j$, where $v_j = s_i$
\STATE $++hash[hreg*hreg\_len+lreg]$
\STATE $state \gets$ `out~of~syllable'
\ELSE
\STATE $++hash[lreg]$
\STATE $lreg \gets s_i - `a\textrm'$
\STATE $state \gets$ `in~syllable'
\ENDIF

\ELSE
\STATE $++hash[lreg]$
\STATE $state \gets$ `out~of~syllable'
\ENDIF

\ENDIF

\IF{$state =$ `in~syllable'}
\STATE $++hash[lreg]$
\ENDIF
\ENDFOR

\STATE \textbf{return} hash
\end{algorithmic}
\end{algorithm}

For easiness of computation and visualization (sacrificing a bit of storage space) the short text is also can be represented in the product superspace $\mathcal{S_{BoSS}} \subset \mathbb{I}^{27} \times \mathbb{I}^{7}$, or a sparse hash of length $27 \times 7 = 189$, where each symbol is calculated as $`0\textrm' + n_{syllable~ occurrences}$ and a bin location calculated as an offset to ASCII symbol `a' and offset to the set \{`a', `i', `u', `e', `o', `y'\} member `a', see Algorithm~\ref{alg:algorithm1}. 

When a new short text comes, the lexical and morphological proximity is calculated as a cosine distance $\cos{\theta}$:

\begin{equation}
\label{eq:1}
\cos{\theta} = \frac{\textbf{v}_1 \cdot \textbf{v}_2}{\|\textbf{v}_1\| \|\textbf{v}_2\|} > t_{\theta}
\end{equation}

and Euclidean distance $d_e$:

\begin{equation}
\label{eq:2}
d_e = {\|\textbf{v}_1 - \textbf{v}_2\|} < t_e
\end{equation}

and compared to the chosen thresholds $t_{\theta}$ and $t_e$, where $\textbf{v}_1$, $\textbf{v}_2$ are short text vector representations in the $\mathcal{S_{BoSS}}$ feature space, see Algorithm~\ref{alg:algorithm2}. C code implementation is publicly available at \url{https://github.com/Selitskiy/BoSS}.

Example texts: ``donald: sprucing up for spring'' and ``vulindlela: sprucing up for spring?''
produce BSS hashes:\\
``0001002000 0102030120 0000000000 0000000000 1000000000 0000001000 

0000000000 0100000000 0100000000 0000000010 0000000000 0000000000 

0000000000 0000000010 1000000000 0000000000 0000000000 0000000000 

000000000''\\ and:\\
``0001002000 0003030120 0000000000 0000000010 0000000000 0000001000 

0000010000 0100000000 0100000000 0000000010 0010000000 0000000001 

0000000000 0000000000 1000000000 0000000000 0000000000 0000000000 

000000000'',\\ 
with $0.885808$ cosine and $2.828427$ Euclidean distances.

\begin{algorithm}
\caption{The Bag of Synthetic Syllables hash comparison function}
\label{alg:algorithm2}
\textbf{Input}: BoSS hashes: $h_1$, $h_2$\\
\textbf{Parameters}: BoSS hash cosine threshold $t_{\theta}$, Euclidean threshold $t_e$\\
\textbf{Output}: BoSS hash proximity flag\\
\begin{algorithmic}[1] 

\FORALL {symbols $s_i$, $i \in \{1, \dots |h_1|\}$}
\STATE $prod \gets prod + (h1_i-`0\textrm') \times (h2_i-`0\textrm')$
\STATE $n2_1 \gets n2_1 + (h1_i-`0\textrm') \times (h1_i-`0\textrm')$
\STATE $n2_2 \gets n2_2 + (h2_i-`0\textrm') \times (h2_i-`0\textrm')$
\STATE $e\_dist2 \gets e\_dist2 + (h1_i-h2_i)^2$
\ENDFOR

\STATE $c\_dist2 = prod^2/(n2_1 \times n2_2)$

\IF{$c\_dist2>t_{\theta}^2 \wedge e\_dist2<t_e^2$}
\STATE $flag=True$
\ELSE
\STATE $flag=False$
\ENDIF

\STATE \textbf{return} flag
\end{algorithmic}
\end{algorithm}

\section{Experiments}
\label{sec:ex}
Experiments were run in the live environment on the Linux Red Hat 7.8 box with 32 GB RAM and Xeon E5-2620 CPU. The BoSS subject header proximity flags were used to generate bulk mail verdicts. Those verdicts, along with the soft SMTP RFC (Request for Comments) standards violations, authenticity verification protocol violations (DKIM \cite{Hansen2011Sep}, SPF \cite{Kitterman2014Apr}, DMARC \cite{Kucherawy2015Mar}, FcRND \cite{Howard2018Nov}), associated DNS record malformity, and traffic pattern artefacts verdicts (overall up to $100$) were fed into a single perceptron classifier. The classifier performed in the near-real-time ($4-5$ million messages per day, or $0.02$ seconds per message processing ) and near-zero foot-print (additional in-memory buffer of the frequent headers of the size $1000$ by $200$ bytes line length) mode that does not require any additional hardware or software enhancement of the SMTP server boxes. The classifier was trained in the reinforcement learning style, where each estimate was used as training data for the next time cycle. The training was done in the semi-supervised mode, in which both crowd-sourced labels and few high-fidelity verdicts were used to form the final training label being in the set $\{spam, not~spam, unknown\}$. Hyper-parameters of the model were set based on the expert estimate and customer feedback, balancing acceptable false positive and false negative error rates.
The processing and storage resources constraints put a limitation on the results collected, especially in terms of comparison with other possible algorithms.

\begin{figure}
\begin{minipage}[b]{0.8\linewidth}
  \centering
  \centerline{
  \includegraphics[width=0.4\linewidth]{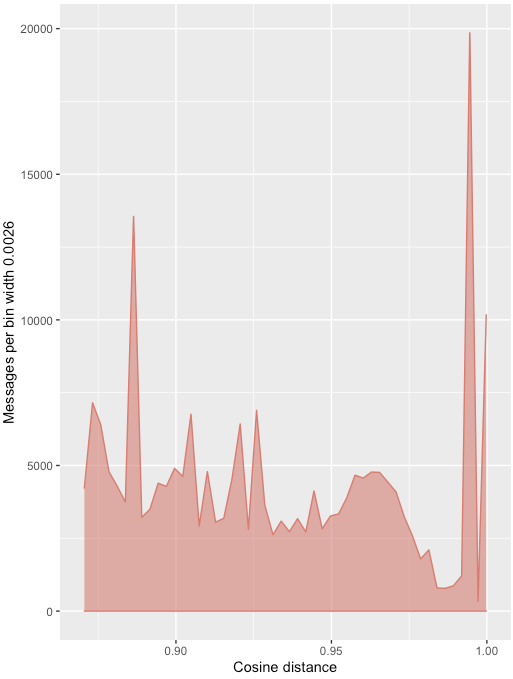}
  \includegraphics[width=0.4\linewidth]{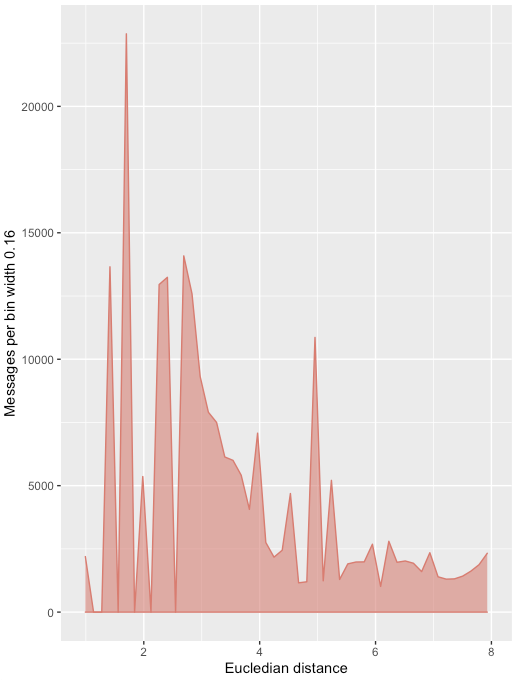}
  }
\end{minipage}
\caption{Cosine (left) and Euclidean (right) distance distribution for messages that triggered BSS proximity verdict. 24 hours snapshot.}
\label{fig:c_dist}
\end{figure}

\begin{figure}
\begin{minipage}[b]{0.8\linewidth}
  \centering
  \centerline{
  \includegraphics[width=0.41\linewidth]{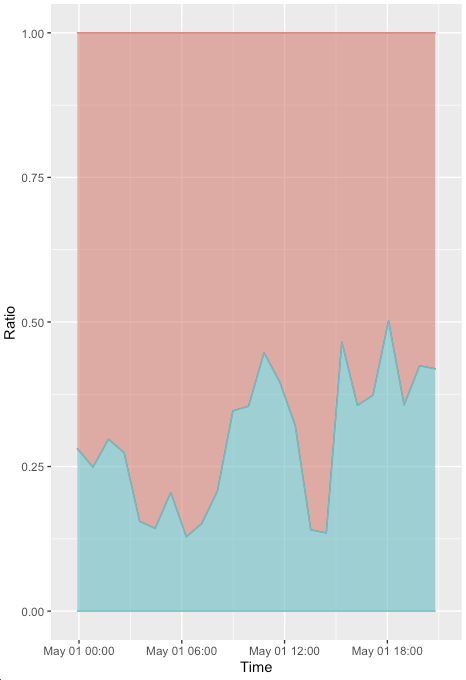}
  \includegraphics[width=0.39\linewidth]{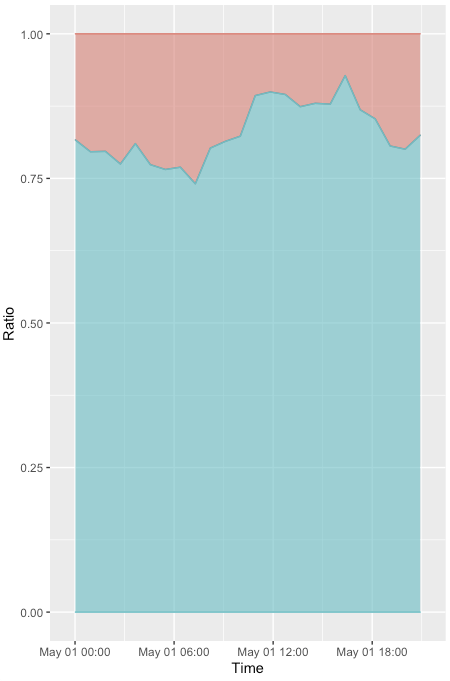}
  }
\end{minipage}
\caption{Proportion of the messages (spam on the left, non-spam on the right) with BoSS spam proximity verdicts (top), and without BoSS spam proximity verdicts (bottom). 24 hours snapshot.}
\label{fig:spam_b}
\end{figure}

\begin{figure}
\begin{minipage}[b]{0.8\linewidth}
  \centering
  \centerline{
  \includegraphics[width=0.4\linewidth]{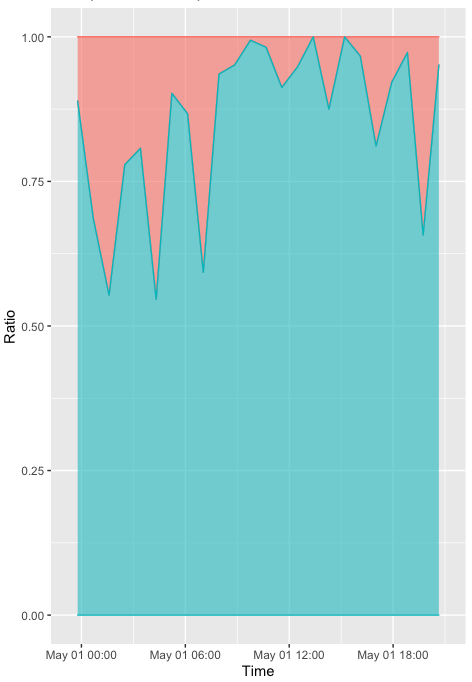}
  \includegraphics[width=0.4\linewidth]{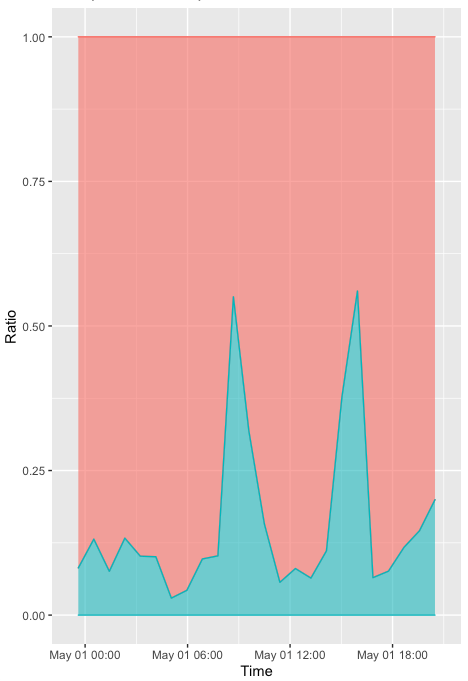}
  }
\end{minipage}
\caption{Proportion of the messages (from banks on the left, social on the right) with BoSS spam proximity verdicts, and without BoSS spam proximity verdicts (bottom). 24 hours snapshot.}
\label{fig:bank_soc}
\end{figure}

Based on the one-day traffic, it can be seen that majority (again circa 90\%) of the bulk mail subject lines is not varying ($2161679$), while $216877$ messages have intentionally or unintentionally mutated subject lines with cosine distance in $(0.87-1.00)$ interval, which was used as a criterion of the subject line variational and morphological proximity for the cosine distance distribution and Figures~\ref{fig:c_dist} for the Euclidean distance distribution. The threshold cosine distance $0.87$ was selected from $\pm0.05$ interval based on the expert estimate and customer feedback.

BoSS proximity verdict associated with particular IP ranges, sender domains, and other source information do not necessarily mean that the messages coming from these sources are spammy but rather indicate the incoming stream's bulk nature.  The bulk mail can be either genuine reporting such as retail or bank statements or social media or subscription notifications that some users may desire and are better to be categorized as grey mail Figures~\ref{fig:bank_soc}.

Therefore, BoSS proximity verdicts are meant to be used with other mentioned above verdicts as input for Machine Learning (ML) algorithms, preferably fast and effective shallow solutions that can utilize the light-weight BoSS approach. Nevertheless, the association of the BoSS proximity verdict with spam verdicts can be seen on Figures~\ref{fig:spam_b}.

\section{Discussion and future work}
\label{sec:disc}

Bag of Synthetic Syllables algorithm offers a less dimensional space than typical Bag of Words algorithms. However, the BoSS algorithm still has enough discriminating power to strongly associate its verdicts with bulk spam or grey mail. Economical, near zero-footprint use of hardware resources and fast near real-time operation allows it to be used as the first line of defence, unloading more sophisticating but slow and resource-demanding DL algorithms.

For future work, a multi-perceptron ML layer working with BoSS verdicts as inputs can distinguish the bad spam verdicts from various flavours of the grey bulk verdicts.

%
%
%








\bibliography{ref.bib}

\begin{thebibliography}{10}
\providecommand{\url}[1]{\texttt{#1}}
\providecommand{\urlprefix}{URL }

\bibitem{Mukti2019}
Akter, M., Jakaite, L.: Extraction of texture features from x-ray images: Case
  of osteoarthritis detection. In: Yang, X.S., Sherratt, S., Dey, N., Joshi, A.
  (eds.) Third International Congress on Information and Communication
  Technology. pp. 143--150. Springer (2019)

\bibitem{BojanowskiGJM16}
Bojanowski, P., Grave, E., Joulin, A., Mikolov, T.: Enriching word vectors with
  subword information. CoRR  abs/1607.04606 (2016)

\bibitem{Hansen2011Sep}
Hansen, T., Kucherawy, M., Crocker, D.: {DomainKeys Identified Mail (DKIM)
  Signatures} (Sep 2011), \url{https://tools.ietf.org/html/rfc6376}, [Online;
  accessed 31. Jan. 2021]

\bibitem{Howard2018Nov}
Howard, L.: {Reverse DNS in IPv6 for Internet Service Providers} (Nov 2018),
  \url{https://tools.ietf.org/html/rfc8501}, [Online; accessed 31. Jan. 2021]

\bibitem{Jain2019}
Jain, G., Sharma, M., Agarwal, B.: Optimizing semantic lstm for spam detection.
  International Journal of Information Technology  11(2),  239--250 (2019)

\bibitem{Jakaite_Balt}
Jakaite, L., Schetinin, V.: Feature selection for bayesian evaluation of trauma
  death risk. In: $14^{th}$ Nordic-Baltic Conference on Biomedical Engineering
  and Medical Physics: NBC 2008 Riga, Latvia. pp. 123--126. Springer Berlin
  Heidelberg (2008)

\bibitem{Jakaite2021}
Jakaite, L., Schetinin, V., Hladuvka, J., Minaev, S., Ambia, A., Krzanowski,
  W.: Deep learning for early detection of pathological changes in x-ray bone
  microstructures: case of osteoarthritis. Scientific Reports  11 (2021)

\bibitem{Jakaite_CMMM}
Jakaite, L., Schetinin, V., Maple, C.: {B}ayesian assessment of newborn brain
  maturity from two-channel sleep electroencephalograms. Computational and
  Mathematical Methods in Medicine pp. 1--7 (2012)

\bibitem{Jakaite2010a}
Jakaite, L., Schetinin, V., Maple, C., Schult, J.: Bayesian decision trees for
  {EEG} assessment of newborn brain maturity. In: The 10th Annual Workshop on
  Computational Intelligence (2010)

\bibitem{Jakaite_EEG_CBMS}
Jakaite, L., Schetinin, V., Schult, J.: Feature extraction from
  electroencephalograms for {B}ayesian assessment of newborn brain maturity.
  In: 24th International Symposium on Computer-Based Medical Systems (CBMS).
  pp. 1--6. Bristol (2011)

\bibitem{Kitterman2014Apr}
Kitterman, S.: {Sender Policy Framework (SPF) for Authorizing Use of Domains in
  Email, Version 1} (Apr 2014), \url{https://tools.ietf.org/html/rfc7208},
  [Online; accessed 31. Jan. 2021]

\bibitem{Klensin2008Oct}
Klensin, J.C.: {Simple Mail Transfer Protocol} (Oct 2008),
  \url{https://tools.ietf.org/html/rfc5321}, [Online; accessed 30. Jan. 2021]

\bibitem{Kucherawy2015Mar}
Kucherawy, M., Zwicky, E.: {Domain-based Message Authentication, Reporting, and
  Conformance (DMARC)} (Mar 2015), \url{https://tools.ietf.org/html/rfc7489},
  [Online; accessed 31. Jan. 2021]

\bibitem{Nyah_IS}
Nyah, N., Jakaite, L., Schetinin, V., Sant, P., Aggoun, A.: Evolving polynomial
  neural networks for detecting abnormal patterns. In: 2016 IEEE 8th
  International Conference on Intelligent Systems. pp. 74--80 (2016)

\bibitem{Nyah_SAI}
Nyah, N., Jakaite, L., Schetinin, V., Sant, P., Aggoun, A.: Learning polynomial
  neural networks of a near-optimal connectivity for detecting abnormal
  patterns in biometric data. In: 2016 SAI Computing Conference. pp. 409--413
  (2016)

\bibitem{Resnick2008Oct}
Resnick, P.W.: {Internet Message Format} (Oct 2008),
  \url{https://tools.ietf.org/html/rfc5322}, [Online; accessed 30. Jan. 2021]

\bibitem{Schetinin2012_ESWA_EEG}
Schetinin, V., Jakaite, L.: Classification of newborn {EEG} maturity with
  {B}ayesian averaging over decision trees. Expert Systems with Applications
  39(10),  9340--9347 (2012)

\bibitem{ijmi2018}
Schetinin, V., Jakaite, L., Krzanowski, W.: Bayesian averaging over decision
  tree models: An application for estimating uncertainty in trauma severity
  scoring. International Journal of Medical Informatics  112,  6 -- 14 (2018)

\bibitem{AIM2018}
Schetinin, V., Jakaite, L., Krzanowski, W.: Bayesian averaging over decision
  tree models for trauma severity scoring. Artificial Intelligence in Medicine
  84,  139--145 (2018)

\bibitem{icae2018}
Schetinin, V., Jakaite, L., Krzanowski, W.: Bayesian learning of models for
  estimating uncertainty in alert systems: Application to air traffic conflict
  avoidance. Integrated Computer-Aided Engineering  26,  1--17 (2018)

\bibitem{eswa_ntdb_2013}
Schetinin, V., Jakaite, L., Krzanowski, W.J.: Prediction of survival
  probabilities with {B}ayesian decision trees. Expert Systems with
  Applications  40(14),  5466 -- 5476 (2013)

\bibitem{Schetinin_EEG_CBMS}
{Schetinin}, V., {Jakaite}, L., {Schult}, J.: Informativeness of sleep cycle
  features in bayesian assessment of newborn electroencephalographic
  maturation. In: 24th International Symposium on Computer-Based Medical
  Systems. pp. 1--6 (2011)

\bibitem{Sureka2010}
Sureka, A., Jalote, P.: Detecting duplicate bug report using character
  n-gram-based features. In: 2010 Asia Pacific Software Engineering Conference.
  pp. 366--374 (2010)

\bibitem{10.1145/3014812.3014815}
Wu, T., Liu, S., Zhang, J., Xiang, Y.: Twitter spam detection based on deep
  learning. In: Proceedings of the Australasian Computer Science Week
  Multiconference. ACSW '17, Association for Computing Machinery, New York, NY,
  USA (2017)

\bibitem{Zhang2010}
Zhang, Y., Jin, R., Zhou, Z.H.: Understanding bag-of-words model: a statistical
  framework. International Journal of Machine Learning and Cybernetics  1(1),
  43--52 (2010)

\end{thebibliography}
\bibliographystyle{splncs03.bst}

\end{document}